\documentclass[%
 print,
 amsmath,amssymb,
 aps,
]{revtex4-2}

\usepackage{graphicx}% Include figure files
\usepackage{dcolumn}% Align table columns on decimal point
\usepackage{bm}% bold math
\usepackage{siunitx}
\usepackage{physics}
\usepackage{xcolor}
\begin{document}

\preprint{APS/123-QED}

\title{Fast and Robust Speckle Pattern Authentication by Scale Invariant Feature Transform algorithm in Physical Unclonable Functions}% Force line breaks with \\
\author{Giuseppe Emanuele Lio}
\altaffiliation{Istituto di Nanoscienze CNR-NANO, Consiglio Nazionale delle Ricerche, Pisa, 56127, Italy}
\author{Mauro Daniel Luigi Bruno}
\altaffiliation{Istituto di Nanotecnologia CNR-NANOTEC, Consiglio Nazionale delle Ricerche, Rende, 87036, Italy}
\author{Francesco Riboli}
\altaffiliation{CNR-INO - National Institute of Optics, Sesto Fiorentino, 50019, (FI), Italy}
\altaffiliation[Also at ]{European Laboratory for Non-Linear Spectroscopy (LENS), University of Florence, Sesto Fiorentino, 50019, (FI), Italy}
\author{Sara Nocentini}
\altaffiliation{Istituto Nazionale di Ricerca Metrologica (INRiM), (TO), 10135, Italy}
\altaffiliation[Also at ]{European Laboratory for Non-Linear Spectroscopy (LENS), University of Florence, Sesto 
Fiorentino, 50019, (FI), Italy}
\email{s.nocentini@inrim.it}
\author{Antonio Ferraro}
\altaffiliation{Istituto di Nanotecnologia CNR-NANOTEC, Consiglio Nazionale delle Ricerche, Rende, 87036, Italy}
\email{antonio.ferraro@cnr.it}

\date{\today}% It is always \today, today,

\begin{abstract}
Nowadays, due to the growing phenomenon of forgery in many fields, the interest in developing new anti-counterfeiting device and cryptography keys, based on the Physical Unclonable Functions (PUFs) paradigm, is widely increased. PUFs are physical hardware with an intrinsic, irreproducible disorder that allows for on-demand cryptographic key extraction. Among them, optical PUF are characterized by a large number of degrees of freedom resulting in higher security and higher sensitivity to environmental conditions. While these promising features led to the growth of advanced fabrication strategies and materials for new PUF devices, their combination with robust recognition algorithm remains largely unexplored.
In this work, we present a metric-independent authentication approach that leverages the Scale Invariant Feature Transform (SIFT) algorithm to extract unique and invariant features from the speckle patterns generated by optical Physical Unclonable Functions (PUFs). 
The application of SIFT to the challenge response pairs (CRPs) protocol allows us to correctly authenticate a client while denying any other fraudulent access. In this way, the authentication process is highly reliable even in presence of response rotation, zooming, and cropping that may occur in consecutive PUF interrogations and to which other postprocessing algorithm are highly sensitive. This characteristics together with the speed of the method  (tens of microseconds for each operation) broaden the applicability and reliability of PUF to practical high-security authentication or merchandise anti-counterfeiting.
\end{abstract}

\maketitle

%\tableofcontents

\section{Introduction}
Coherent light that impinges and diffuses into a scattering medium, interferes in the far field producing a granular image referred to as speckle pattern \cite{goodman1976some,dainty2013laser}. The speckle is both a challenge and an opportunity in many fields, as it can degrade image quality or provide useful information for measuring surface roughness, displacement, and biological parameters finding therefore applications in various scientific fields spanning from astronomy, imaging, cultural heritage, metrology and crypto-security \cite{dong2017review,basak2012review,senarathna2013laser,pino2010method,farid2023speckle,tornari2019development,pappu2002physical, kim2022revisiting, cao2022harnessing}. 
In the latter, speckle patterns originated by illuminating a rough or light-scattering surface can act as a unique “fingerprint” that is nearly impossible to replicate generating the so called optical physical unclonable functions (PUFs). Due to the possible multiple interrogations and large number of degrees of freedom, PUFs work as on-the-fly generators of secure cryptographic keys \cite{pappu2002physical,mcgrath2019puf,wan2021bionic,ferraro2023hybrid,nocentini2024all}. Due to their inherent manufacturing errors and internal randomness, it has been demonstrated that reproduction of the same optical PUF is impossible even by the manufacturer itself \cite{bin2020robust,lio2023quantifying}. Optical strong PUFs are typically used within a Challenge Response Pairs (CRPs) protocol that includes an enrollment and verification process.
In both steps, the scattering sample is illuminated with multiple pseudo-random challenges ($C$) and the corresponding responses ($R$) are collected by a camera. During the enrollment process, a large database of CRPs is stored at the central authority as reference and the PUF is then delivered to the client.\\
The prover authentication is then based on the analysis of speckle pattern images that, at first, are post-processed and then analyzed by statistical methods where parameters such as contrast, correlation length, and intensity distribution are used to quantify speckle characteristics \cite{pappu2002physical, hussain2018shaip, usmani2018efficient}.  
Post-processing can be performed using standard image transformation (e.g. the Gabor hashing or wavelet decomposition) and binarization algorithms. Along this process, the wavelength of a wavelet-based Gabor filter is tuned to extract the relevant features of the speckle images, both ensuring the repeatability of the responses under the same challenge interrogation and preserving the random nature of responses to different challenges. The response $R_i$ (speckles) related to the challenges $C_i$ is then hashed and reshaped into a 1D binary array or keys $K_i$. The pairwise distance between each binary keys $K_1, \dots, K_i$ is then measured with the Hamming distance metric, i.e. the number of bits that differ in two bit strings\cite{daugman2003importance}. Distances between keys generated by different challenges are called ``unlike'' distances (and are related to the entropy of the key), while those generated by same challenges are called ``like'' distances (and are related to the stability of the PUF).  Commonly, the Fractional Hamming Distance (FHD) metric, the Hamming distance normalized by the bit string length, is also used to compare the binary keys that are retrieved from the speckle patterns recorded during the authentication with the ones of the database collected during the enrollment. Moreover, the currently employed analysis approaches suffer from a high sensitivity to minimal variations of the speckle pattern that may occur in case of minimal variation of the position, illumination and rotations that may lead to a failure in the PUF authentication\cite{lio2023quantifying}.
\\In this scenario, it is of paramount importance to find a new, flexible analysis method for fast processing of speckle pattern. Image recognition algorithms enables the analysis of images to extract relevant information and features. These algorithms are being improved at fast pace and are now used in daily life in many sector such us logistic, security access, healthcare, traffic management and so on \cite{russakovsky2015imagenet}. 
Scale-Invariant Feature Transform (SIFT) \cite{Lowe1999obj,Lowe2004dist} represents a valid and versatile method for detecting and extracting distinctive local features from an image \cite{bicego2006use, badrinath2008palmprint}. A key characteristic is the ability to recognize unique features of the image under analysis also under transformations such as scaling, rotation and  illumination variations. 
This robustness has made SIFT widely utilized in computer vision tasks, particularly in object recognition and image matching \cite{burger2022scale}. A decade ago, a preliminary work using SIFT for speckle recognition shown how it improves the accuracy identification with respect to correlation analysis in tens of seconds \cite{yeh2012robust}.
Based on this study, more recently, we leverage on its capability of extracting unique and stable features for anti-counterfeiting applications. Specifically, SIFT enables the identification and authentication of security tags by detecting keypoints, comparable to a fingerprint. These features serve as digital signatures, which can be stored and subsequently compared against a reference database for verification purposes \cite{ferraro2022low,bruno2023cholesteric,kamwe2024optical}.
It is remarkable that for human fingerprints, a minimum number of 20 unique features matches is enough to agree with the identity of the designed subject \cite{champod2013fingerprints}. 
\\In this work, we present a straightforward approach based on Scale-Invariant Feature Transform (SIFT) algorithm for rapid and robust identification and validation of strong optical PUFs whose speckle patterns generated through the Challenge-Response Pair (CRP) protocol. 
To demonstrate the effectiveness of our approach, we fabricated three distinct types of optical PUFs, each characterized by different scattering properties. The first type is fabricated by dispersing polystyrene (PS) nanoparticles (NPs) onto a glass substrate. The second involved the realization of a polymer-dispersed liquid crystal (PDLC) structure, while the third is based on titanium dioxide (TiO$_2$) nanoparticles embedded within a polymer matrix. We refer to these implementations as PS-PUF, PDLC-PUF, and TiO$_2$-PUF, respectively.
The produced speckle patterns were acquired, stored into different datasets and processed on-demand through the SIFT algorithm. To prove the reliability of the proposed method, speckle patterns underwent distortion such as rotation, scaling and cropping without affection the identification capability of the SIFT.
The ability of SIFT to extract distinctive and stable features from complex speckle patterns underscores its suitability for fast and accurate authentication, paving the way for enhanced security applications in anti-counterfeiting and identification systems.

\section{Materials and Method}
\subsection{Three PUF-Samples and Their Fabrication}
\begin{figure}[!h]
\centering
\includegraphics[width=14cm]{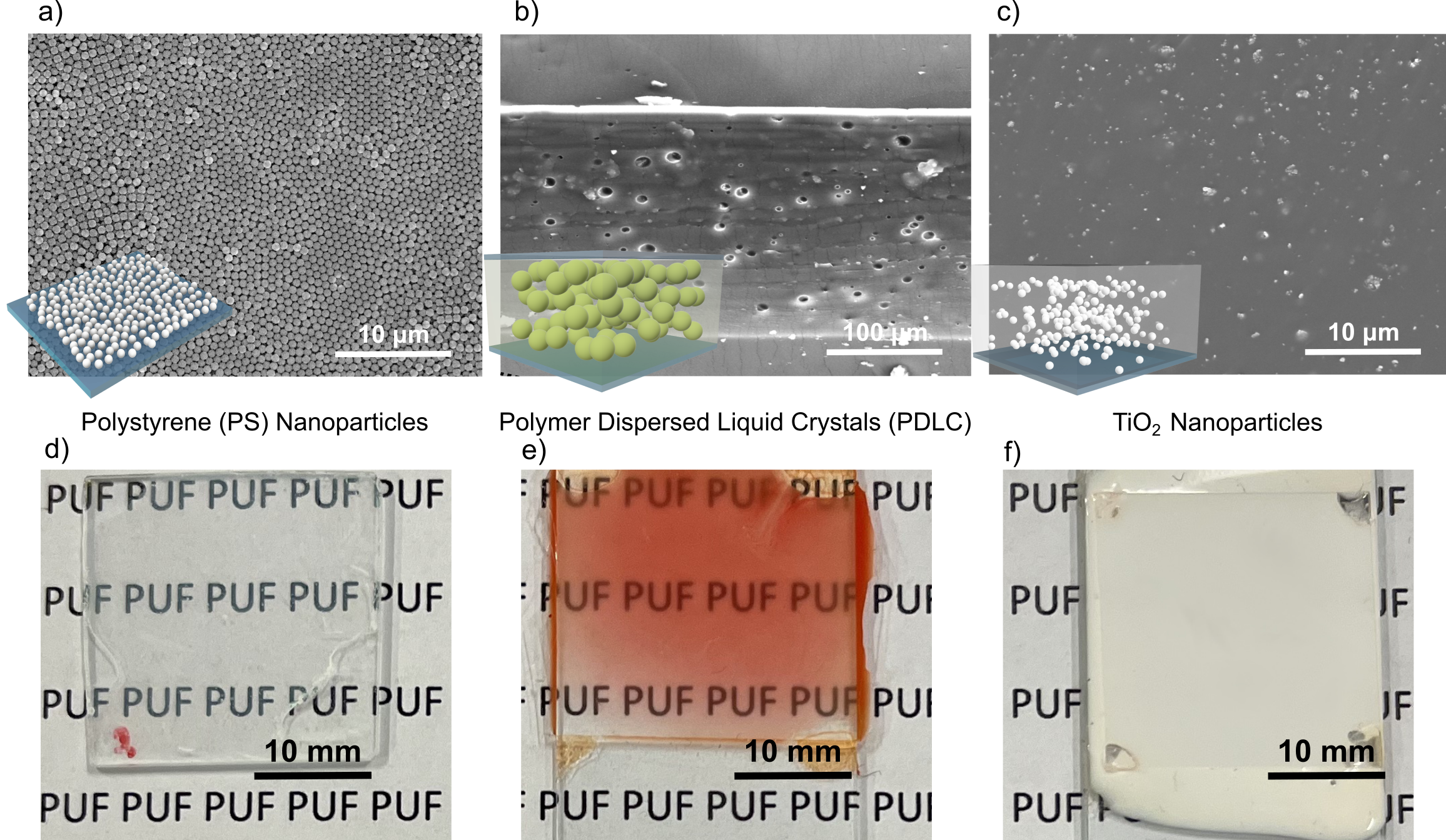}
\caption{SEM images and schematic representation of a) PS-PUF constituted by a single layer of polystyrene nanoparticles, b) PDLC-PUF constituted Polymer Dispersed Liquid Crystals PDLC-PUF and c) TiO$_2$-PUF constituted by TiO$_2$ nanoparticles dispersed in a dense polymer matrix. d-f) photographs of the proposed three PUFs.}
\label{sample}
\end{figure}

We have fabricated and experimentally characterized three PUF-samples. They are made from distinct materials and possess different optical characteristics, such as internal entropy, scatterer size and density, and so-called optical thickness ($OT$), that is the natural logarithm (ln) of the sample transmission. As a simple quantitative example, an optical thickness of 1 indicates that the transmitted/diffused intensity is reduced by a factor of 1/e with respect to the initially incident intensity. Samples with small optical thickness around $OT\leq1$, are amost transparent to a human observer, and the photons traveling through the sample on average undergo only a few scattering events. On the contrary, samples with larger optical thickness $OT>1$ make the light bounce
multiple times internally among their scattering centers and appear increasingly opaque to the human eye. \\
The PS-PUF-sample shown in Figs.\ref{sample}a-d consists of a single layer of polystyrene nanospheres of diameters $d \simeq 250nm$, which were randomly deposited on a glass substrate ($25 \times 25 mm^2$ in x-y-dimensions) by a spin coating process. By tuning the spin coating rotation speed and keeping the other parameters fixed, either a single- or double-layer packing can be obtained. For our purposes, a single layer PUF with thickness around $250nm$ in z-dimension was created by using a rotation speed of 2000 rpm. As desired, the sample presents small crystalline regions with a hexagonal close-packed symmetry, few voids, and no double layers. The low refractive index of polymer nanospheres ($n \simeq 1.59$) as well as the single layer geometry leads to a weakly scattering medium and the resulting refractive index contrast is 0.59. Furtheremore, it is fully transparent to the human observer, with an optical thickness of OT = 0.39.\\
The PDLC-PUF-sample illustrated in Fig\ref{sample}b-e consists of a polymer-dispersed liquid crystal (PDLC) structure, with x-y-dimensions of again $25 \times 25 mm^2$ and a thickness in z-direction of $140\mu m$. The sample is obtained by an emulsion of liquid crystal molecules (5CB) into a matrix of polydimethylsiloxane (PDMS). The liquid crystals form light scattering droplets with a diameter of around $10\mu m$. Please note that in opposition to our first PS-PUF-sample, we are no longer forming a single layer in z-dimension, whose thickness is roughly equal to the diameter of its scattering elements. Instead, the thickness of the second PUF-sample is about 15 times the size of its scatterers. The small refractive index contrast between the scatters (droplets made by radially aligned liquid crystal molecules with average refractive index $n_{av}\simeq 1.6$ at $25^o C$ and wavelength $\lambda \simeq 633 nm$) and the polymeric matrix ($n\simeq 1.43$) results in a weakly scattering medium, refractive index contrast of 0.17 and optical thickness less than 1 (OT = 0.7). It exhibits a slightly reduced transparency to the human eye.\\
Finally, the $TiO_2$ PUF-sample shown in Fig \ref{sample}c-f consists of Titanium dioxide ($TiO_2$) nanoparticles in a polymeric stabilizing matrix, more precisely in a photopolymer resin. It is made of a commercial UV-curing acrylate optical adhesive with a dispersion of rutile $TiO_2$ nanoparticles with a diameter of $280 nm$. The mixture of polymer and nanoparticles is rendered homogeneous through magnetic stirring and an ultrasonic bath for around 1 hour. It is then cured with an UV lamp, resulting in a $230\mu m$ thick film on a glass substrate. The large refractive index of $TiO_2$ ($n \simeq 2.87$) compared to the surrounding polymeric matrix ($n \simeq 1.52$) leads to a strong scattering with a refractive index contrast of 1.35. It creates an optical thickness of $OT = 2.25$, and a structure that is opaque to human observers. Table \ref{table_sample} summarizes the properties of the three PUF-samples.\\

\begin{table}[hb]
\begin{tabular}{lccccc}
\multicolumn{1}{c}{\textbf{PUF}} & \textbf{Scatterer size}      & \textbf{Number of Scatterers} & \textbf{Scatterer Density} & \textbf{Optical Thickness} & \textbf{R.I. Contrast} \\
PS                               & d $\sim$ 250 nm & 10$^2$                        & 10$^9$/cm$^2$              & 0.39                       & 0.59                  \\
PDLC                             & d $\sim$ 10 µm  & 10$^6$                        & 7*10$^7$/cm$^3$            & 0.7                        & 0.17                   \\
TiO2                             & d $\sim$ 250 nm & 5*10$^{10}$                     & 2*10$^12$/cm$^3$           & 2.25                       & 1.35                
\end{tabular}
\caption{Properties of the three optical PUFS}
\label{table_sample}
\end{table}

%%%%%%%%%%%%%%%%%%%%%%%%%%%%%%%
%%%%%%%%%%%%%%%%%%%%%%%%%%%%%%%

\subsection{Challenge Response Pair Protocol and optical apparatus}
\begin{figure}[!h]
\centering
\includegraphics[width=14cm]{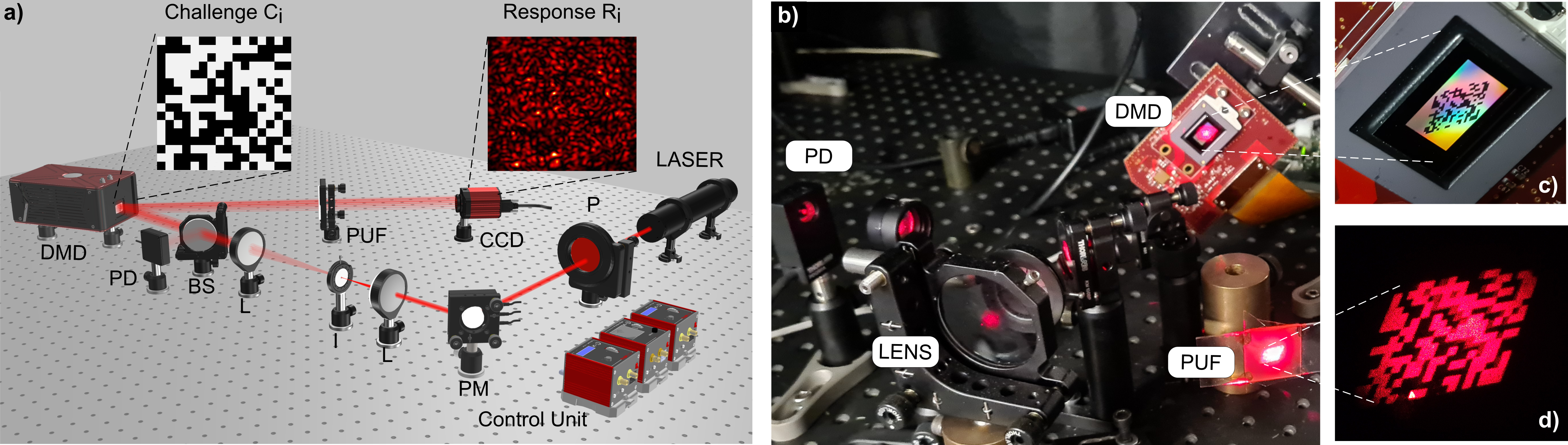}
\caption{a) Schematic representation of the experimental setup used to collect the CRPs. b) Real picture of the main part of the setup used to generate and project the challenges. c) The challenge on the DMD, d) the projected challenge on the PUFs.}
\label{setup}
\end{figure}

The optical PUFs were investigated using Challenge Response Pairs (CRPs) protocol \cite{pappu2002physical, bruno2024flexible,ferraro2023hybrid,lio2023quantifying}. A He-Ne red laser beam with wavelength $\lambda=633\:nm$ (power $5$ mW) propagates through a series of lenses, polarizers, and irises. The beam, after a beam-expander, impinges on a digital micro-mirror device (DMD) used for the spatial intensity modulation of the laser beam that is then conveying the challenge ($C_i$) to the PUF. Each individual $C_i$ is then projected on the PUF surface and light diffuses through the disordered media. The transmitted light interferes in the far field producing an optical pattern named speckle pattern. This constitutes the PUF response $R_i$, which is collected in cross-polarization configuration to remove any non-scattered light. This pattern is collected by a CCD camera, Thorlabs camera CS165MU. Here, we used a $270\times360$ px camera with $40$ FPS for this task. \textbf{Figure \ref {setup}} reports a schematic representation and photos of the experimental setup.

%%%%%%%%%%%%%%%%%%%%%%%%%%%%%%%
%%%%%%%%%%%%%%%%%%%%%%%%%%%%%%%

\subsection{Scale Invariant Feature Transform analysis}
The produced speckle patterns were analyzed using SIFT algorithm, which identifies image features by transforming an image into a vast collection of local feature vectors. These latter, identified as SIFT keys, are invariant to translation, scaling, rotation, and partially to illumination changes in the image \cite{Lowe2004dist}, making them more robust also in case of fluctuation of the intensity of the laser beam. Four main calculation steps are required to generate all the image features: (a) scale-space extrema detection, done using a Gaussian difference function to identify potential points of interest that are invariant to scale and orientation; (b) key point localization, where key points are selected based on measurements of their stability; (c) orientation assignment, where one or more orientations are assigned to each key point location based on the gradient directions of the local image; and (d) key point descriptor, which describes key points by measuring the local gradients in the image at the selected scale \cite{Lowe1999obj}. To analyze and compare the images, a script was used implementing the SIFT algorithm, from OpenCV library, in Python \cite{kamwe2024optical}. The SIFT algorithm identifies keypoints and evaluates the typical pixel size using scale-space extrema detection, which relies on differences of Euclidean distances (L2 norm) to detect significant image structures across multiple scales. Then, several parameters are used to configure the feature recognition process, such as the maximum number of features the user wishes to detect, the number of octaves in each Difference-of-Gaussian (DoG) function (with 4 being a commonly recommended value according to OpenCV documentation), and the contrast and edge thresholds, which help in accurately identifying bright and dark features within the image. To construct the first octave, a Gaussian filter is applied to the input image using multiple values of $\sigma$. For the second and subsequent octaves, the image is first downsampled by a factor of 1.6, and then Gaussian filters with different $\sigma$ values are applied. Specifically, Octave 1 uses a scale of $\sigma$, Octave 2 uses 2$\sigma$, and so on. After setting these parameters, a matching distance (Md) is defined. This value quantifies the number of common features detected between two compared images. \cite{wu2013comparative, sima2013optimizing, wang2019approximated}. According to the quality of the collected figures and after optimization procedure, we identified and used for all the analyses the following parameters: the number of recognized features set equal to 0 (it means to find all possible matches) the number of octave layers equal to 4, the contrast and edge threshold equal to 0.04 and 5, respectively. Finally $\sigma$ has been set at 1.6 and Md at 0.7. While these parameters can be finely adjusted based on feature size and image contrast to optimize the number of detected keypoints, we demonstrate that, even with a fixed optical setup, varying the system's optical properties allows the SIFT algorithm to reliably authenticate different primitives.

\section{Results}
%%%%%%%%%%%%%%%%%%%%%%%%%%%%%%%%
\begin{figure}[!h]
\centering
\includegraphics[width=14cm]{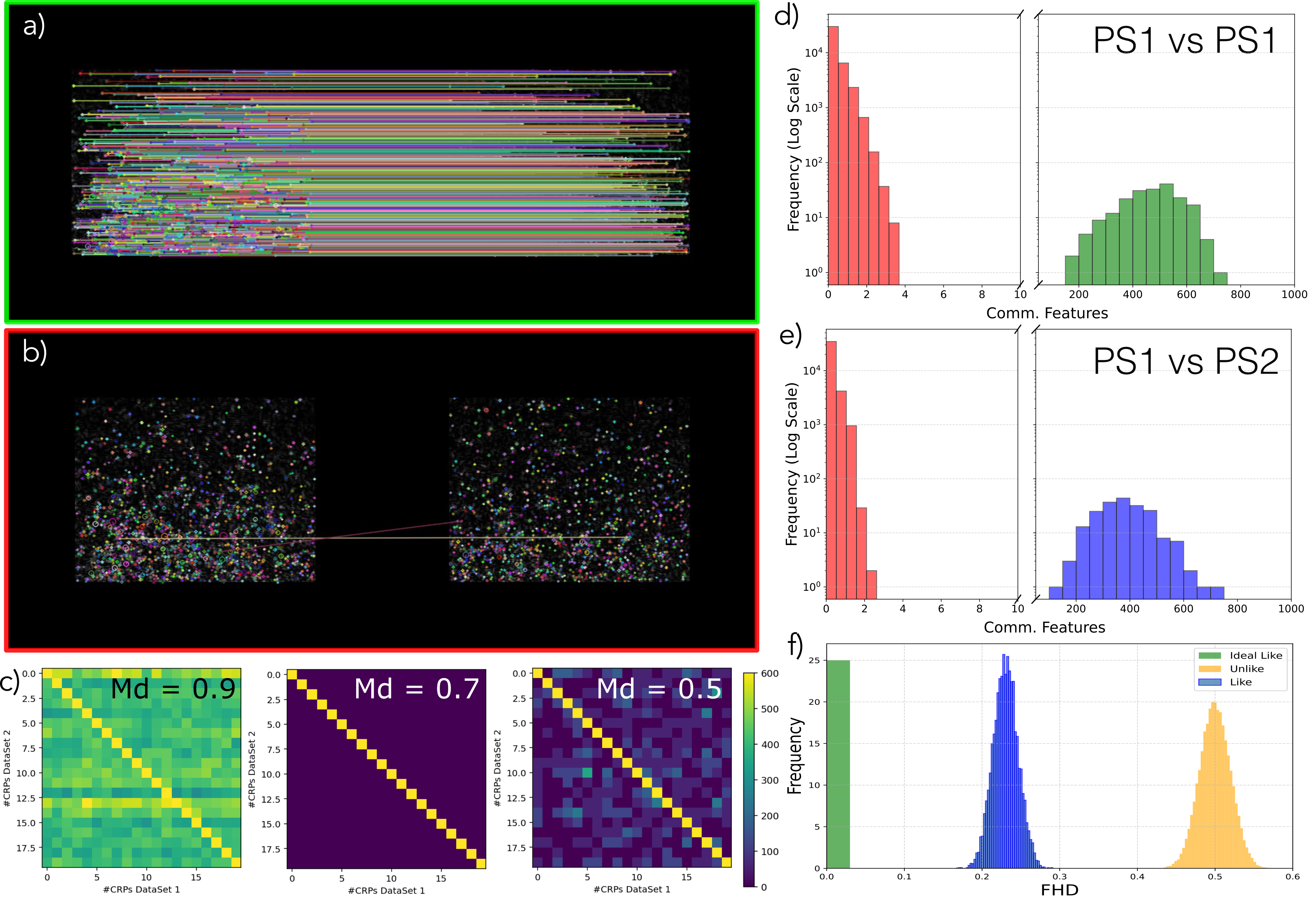}
\caption{a–b) Example images showing fully recognized speckles (produced under the same challenge conditions) and cases with no or few recognized points (from different speckles). c) Maps showing recognized points for a comparison of 20 versus 20 speckles, with varying matching distance (Md) parameter value. The scale is the same for the three matrices with a maximum number of recognized point of 600.  d) Comparison between 200 versus 200 speckles from the same dataset (PS1). In the histogram the speckles that found their match are reported in green while the other one with no match are reported in red.
e) Comparison between 200 versus 200 speckles from two datasets collected at different times (PS1 and PS2). In the histogram the speckles that found their match are reported in blue while the other one with no match are reported in red. f) Fractional Hamming Distance (FHD) distribution calculated over the larger database, illustrating the 'ideal-like,' 'like,' and 'unlike' distributions.}
\label{SL_dataset}
\end{figure}
%%%%%%%%%%%%%%%%%%%%%%%%%%%%%%%

In order to evaluate the SIFT-based authentication on different types of optical PUFs, we fabricated three primitives that produce trasmission speckle patterns with different brightness and contrast. We selected i) a single layer of  Polystyrene (PS) NPs on glass, ii) a 3D low refractive index PUF made by a polymer dispersed liquid crystals (PDLC)  and iii) a 3D high refractive index medium by dispersing Titanium Dioxide (TiO$_2$) NPs into polymer matrix (for further details see Materials and Method section).

We first investigated the properties of PS-PUF by interrogating it with 200 challenges $C_i$ and collecting the resulting response $R_i$ as speckle patterns (see Method section for details). They were then analyzed with the SIFT algorithm and the results reported in \textbf{Figure \ref {SL_dataset}}. We refer to this speckle patterns database as dataset PS1 acquired at time t0. This analysis results in a database of 40000  comparisons. When comparing each speckle pattern with itself, the algorithm recognizes a large number (a few hundreds) of unique feature matches which are visually indicated as colored linked lines as shown in \textbf{Figure \ref {SL_dataset}}a. When the SIFT algorithm is applied to two different responses, the common matched features are few (less than five), namely are false positive, as shown in \textbf{Figure \ref {SL_dataset}}b.
\textbf{Figure \ref {SL_dataset}}c presents three correlation maps of the number of identified features for 20 different speckle patterns varying the matching distance (Md). Herein, the diagonal represents the self-comparison of each speckle pattern with values exceeding 600. The maps show that, for a Md value of 0.9 or 0.5, the number of false-positive recognition increases, while a value of Md of 0.7 allows to correctly recognize the speckles with themselves indicating this parameter as the best choice for the selected patterns. 
\textbf{Figure \ref {SL_dataset}}d reports the histogram of the number of false positive (red bars) and true positive (green bars) matches for the entire comparison of PS1 database where the green bars correspond to speckles that found a number of matches that overcomes 400 points. The variability of the matched features in between identical speckle patterns depends on the number of speckle grains present in the pattern as well as their contrast. The bars indicate also the number of common features and related occurrence frequency. The red bar, instead, reports the number of comparisons with a few matches found, values that remains below ten points for all the considered speckles. 
Next, the same PS-PUF was analyzed at a different time using the same CRPs protocol, with the same challenges set (C$_{i...N}$). The resulting speckle patterns were acquired to form a new dataset, referred to as PS2 acquired at time t1.
Both datasets were then used as a database for the SIFT comparison. As shown in Figure \ref{SL_dataset}e, the number of matched points is maximized only for 200 speckle patterns (blue bars), corresponding to the responses ($R_i(t_0)$ and $R_i(t_1)$) generated using the same challenge ($C_i$). Even if there is a reduction of the common matched features due to environmental fluctuations, it is still possible to clearly recognize each single speckle pattern that have more than 200 common features. The blue histogram, therefore, is now indicating the stability of the system: the larger number of recognized points the higher is the system stability. In contrast, as expected, all other cross-comparisons ($R_i(t_0)$ and $R_j(t_1)$) generated from the challenge ($C_i$ and $C_j$) resulted in only a few recognized matched points (red bar). This confirms the uniqueness and large difference of the responses related to different challenges. As counterproof, we analyzed the speckles pattern datasets (PS1 versus PS2) with the FHD metric, see \textbf{Figure \ref {SL_dataset}}f. By comparing the speckle $R_i(t_0)$ from PS1 with $R_j(t_1)$ from PS2, which are the outputs produced by the same $C_i$, we obtain a "like" distribution (indicated in blue) with a mean value at 0.23. Hence, the blue histogram shown in Figure \ref{SL_dataset}e can be considered the analog of the "like" distribution reported in the FHD analysis. The recognized responses represented in the blue histogram exhibit a shift in the number of common features toward lower values similar to the FHD “like” distribution, that shifts towards larger FHDs with respect to the ideal value (green bar). On the other hand, when comparing speckle responses generated using different input challenges ($C_i$), an “unlike” distribution is obtained, peaked at around 0.5, as expected. This analysis confirms that the speckle patterns produced by different challenges are inherently distinct, thereby demonstrating the capability of the proposed PUF to generate unique, but reproducible responses with no false-positive matches in the SIFT analysis. As with the FHD metric, the clear separation between the “like” and “unlike” distributions in the SIFT approach enables the definition of an authentication threshold that minimizes the false positive. For the PS-PUF, a threshold of approximately 100 common features can be adopted to ensure reliable authentication.
%%
%%%%%%%%%%%%%%%%%%%%%%%%%%%%%%%
\begin{figure}[!h]
\centering
\includegraphics[width=14cm]{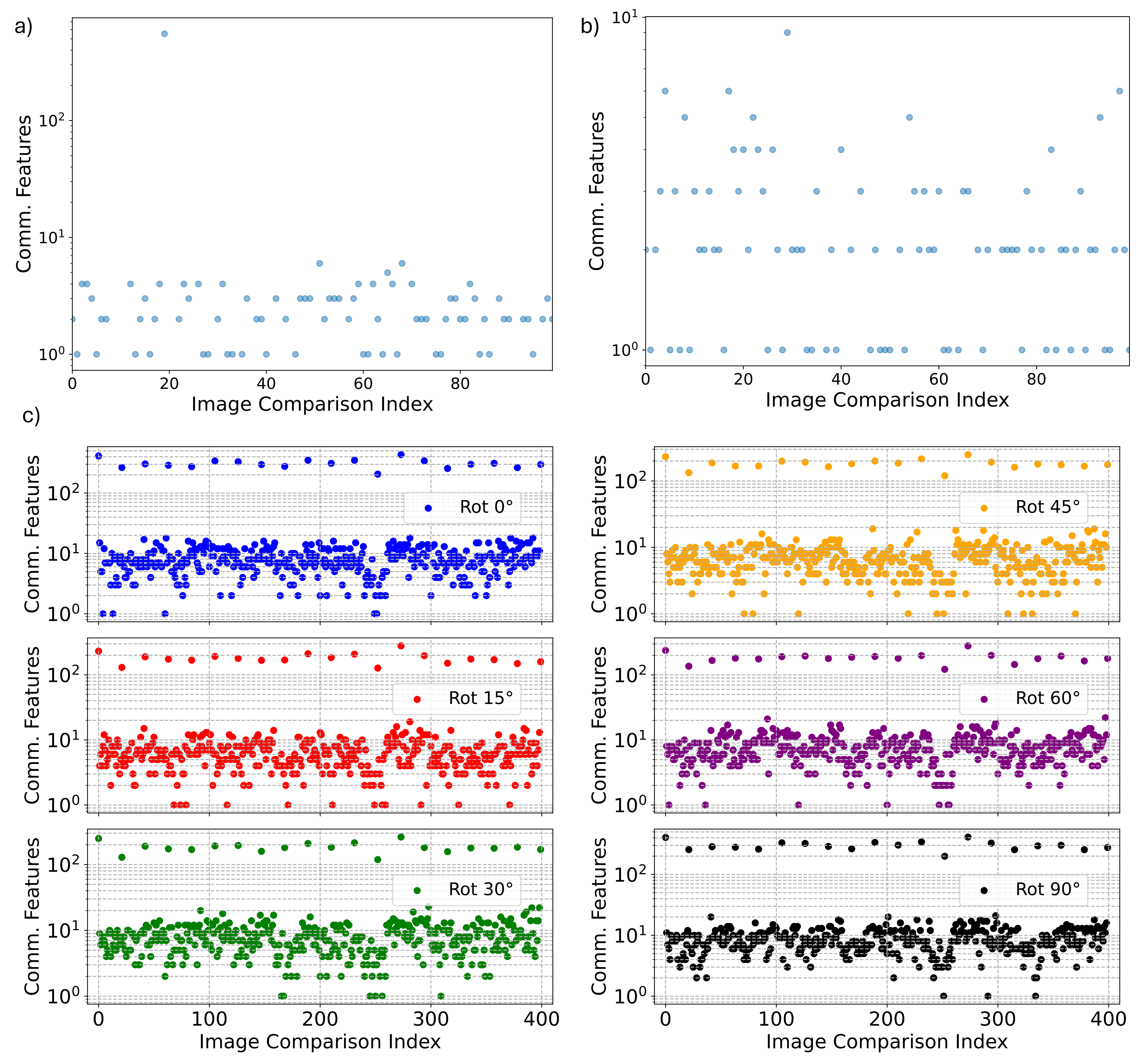}
\caption{Comparison, using SIFT analysis, of 1 speckle into a database of 100 speckles which is a) inside, b) outside the database. c) Comparison of speckle patterns, using SIFT analysis, by rotating them at 0°, 15°, 30°, 45°, 60° and 90°}
\label{1vs100}
\end{figure}
%%%%%%%%%%%%%%%%%%%%%%%%%%%%%%%
Given this characterization, we evaluated the effectiveness of the proposed SIFT-based approach in identifying, and hence authenticating, a single speckle pattern referred to as $Sx$, within a database, in other words the ability to identify if a speckle is part or not of a certain database (PS1 or PS2). In the cloud of points map shown in \textbf{Figure \ref{1vs100}a}, where the X-axis indicates the speckle index and the Y-axis the number of common unique features, there is a clear evidence of a single response for which the number of recognized points overcome the authentication threshold (response \#$20$ having more than 500 common features). In contrast, all other speckles exhibit only a few common features, which can be considered as “false positives.” 
To further validate this approach, we repeated the analysis considering a target speckle $Sy$, e.g the response produced by challenge 150, that is outside the database. The results, presented in \textbf{Figure \ref{1vs100}b}, show that for all speckle patterns (again 100) in the database, only a few “false positive” matches were detected. Please note that on Y-axis the maximum value now is 10. This confirms that when the queried speckle is absent from the database (PS1 or PS2), SIFT does not falsely identify any pattern. These findings demonstrate that once a dataset is established, the proposed SIFT-based method can reliably recognize and verify the presence of a unique speckle pattern. 
One of the strengths of the SIFT method is its invariance to image rotation, a crucial factor when analyzing speckle patterns that are inherently sensitive to even small variations in orientation. In order to confirm such robustness, we analyzed a database of 20 speckle patterns (from PS1) and compared them with themselves after the application of an image rotation (PS1-R) at seven different angles namely 0°, 15°, 30°, 45°, 60° and 90°. The results, shown in Figure \textbf{Figure \ref{1vs100}c}, are presented using waterfall plots to illustrate that only for 20 comparisons of the response with itself after rotation, the commonly matched features overcome 100. In contrast, for all the other possible comparisons, the  number of unrecognized features falls in the range in between 1 and 20. %Notably, per each rotation only 20 points exhibit a high number of matched features (>100), corresponding to the number of speckles in the considered dataset, demonstrating that each speckle pattern is consistently recognized as itself, regardless of rotation. 
This confirms that SIFT effectively preserves its matching accuracy even when speckle patterns undergo significant angular transformations.
%%%%%%%%%%%%%%%%%%%%%%%%%%%%%%%
\begin{figure}[hb]
\centering
\includegraphics[width=9cm]{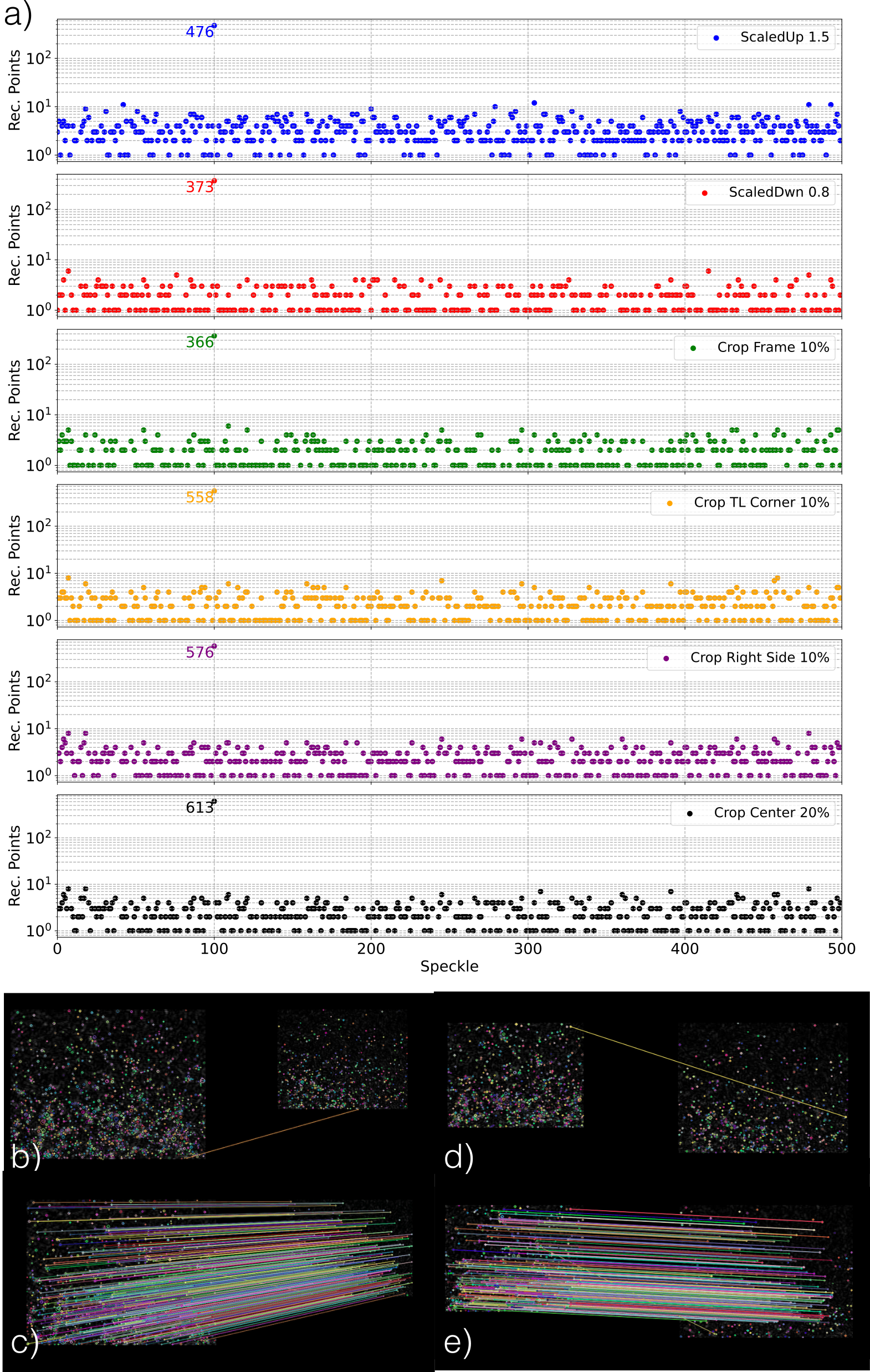}
\caption{a) Comparison, using SIFT analysis, of 1 speckle (tag number $\#$100) into a database of 500 ones when scaled-up by 1.5, scaled-down by 0.8, cropped of 10$\%$ along the frame, cropped of 10$\%$ from the Top-Left (TL) corner, or from the right side and finally cropped of 20$\%$ in the center. Zero and full recognized points from scale-up tag (b,c) and cropping the surrounding frame by 10$\%$ (d-e).}
\label{scale_crop}
\end{figure}
%%%%%%%%%%%%%%%%%%%%%%%%%%%%%%%
In the following analysis, we performed an additional test to corroborate the robustness of the SIFT method applied for the recognition of speckle patterns. In fact, SIFT enables fast and reliable identification of a speckle, even when the pattern is subjected to transformations such as scaling or cropping (from all edges or a single side) that can occur for example when using different optical setups.
This capability is demonstrated in \textbf{Figure \ref{scale_crop}a}, where SIFT successfully detects and verifies the presence of the speckle under consideration despite these modifications.
The SIFT protocol was tested by checking the correct identification of the speckle pattern $\#$100 after different image transformations within a database of 500 speckle patterns. The speckle pattern $\#$100 was subjected to: scaling-up by 1.5, scaling-down by 0.8, cropping of 10$\%$ along the frame, cropping of 10$\%$ from the Top-Left (TL) corner, or from the right side and finally cropping of 20$\%$ in the center. 
The results, illustrated in \textbf{Figure \ref{scale_crop}a}, confirm that even when the target speckle pattern ($\#$100 for this case) undergoes such transformations, there are a large number of matched features, well above the authentication threshold only in the correspondence of speckle $\#$100 inside the database, demonstrating therefore high specificity and accuracy. \textbf{Figure \ref{scale_crop}b–e} shows clear cases of successful and failed recognition. Specifically, it reports zero and full feature matches in two test scenarios: one where the speckle pattern was scaled up by a factor of 1.5 (b,c), and another where it was cropped by 10$\%$ along the tag frame (d,e). These findings reinforce the adaptability of SIFT compared to other method used to identify and grant the originality of PUFs, making it a highly effective tool for authentication under varying imaging conditions.

%%%%%%%%%%%%%%%%%%%%%%%%%%%%%%%
\begin{figure}[!h]
\centering
\includegraphics[width=14cm]{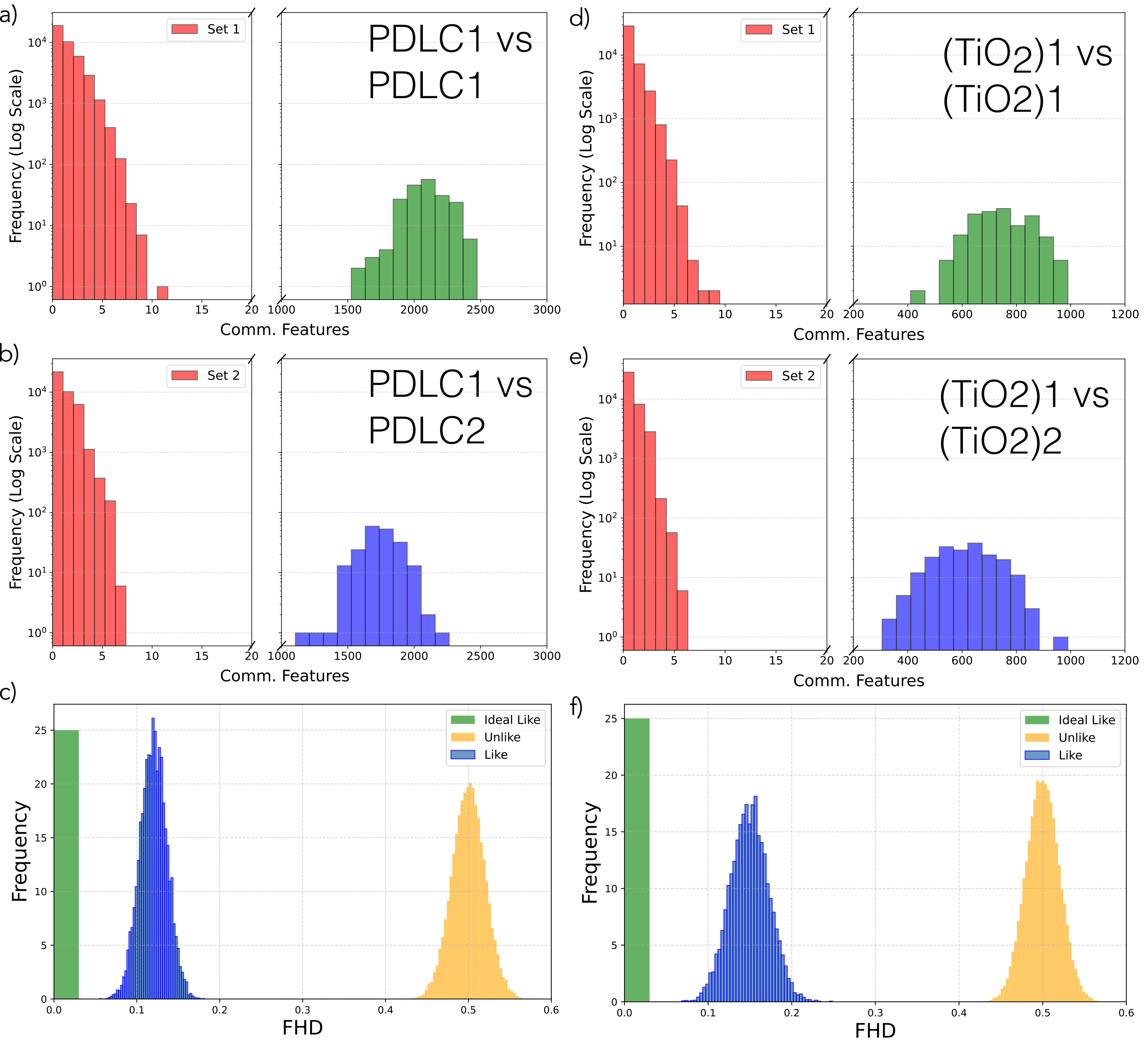}
\caption{SIFT applied to CRPs from the same dataset (PDLC1, (TiO2)1) and between two dataset collected at two different time (PDLC2, (TiO2)2) on a,b) the PDLC and d,e) the TiO$_2$ PUF. In the histogram the speckles that found their match are reported in green (same dataset) and in blue (for different dataset) while the other with no match are reported in red in both cases. c,f) Report the FHD for both PUFs typologies displaying the "ideal like" the analogue of SIFT green histogram, "like" (the analogue of SIFT blue histogram) and "unlike" distributions.}
\label{PDLCTiO2}
\end{figure}
%%%%%%%%%%%%%%%%%%%%%%%%%%%%%%% 
To further demonstrate SIFT ability to analyze speckle patterns generated by different types of optical PUFs, possessing different refractive index contrast and scattering strengths, we applied both the SIFT (using the same SIFT parameters) and FHD methods to PDLC and TiO$_2$ PUFs. Also for this analysis, a representative dataset of 200 speckles was used. We refer to this as PDLC1 and (TiO2)1 respectively. As reported in \textbf{Figure \ref {PDLCTiO2}}a,d, for both typologies, we obtained the same behavior of PS-PUF previously tested. However, it is worth noting that for the PDLC-PUF the number of matched recognized points is more than doubled with respect to PS-PUF ranging, now, between 1500- 2500 points. This is because more scattering particles (LC molecules for this PUF) are present in the volume of the analyzed PUF, producing more well-defined speckle grains into the speckle patterns, hence increasing the number of features that can be recognized and matched in the algorithm execution. 
In contrast, for the TiO$_2$-PUF, the number of matched recognition points is approximately 800–1000, despite the high density of scatterers. This reduced number is attributed to the low transparency of the PUF having an OT=2.25, which lowers the contrast of the speckle grains in the collected images. Nevertheless, even with fewer points and reduced contrast, the SIFT based authentication is validated also for strongly scattering systems without the need of a fine tuning of the algorithm parameters. We can thus apply this methodology to a wide variety of optical PUF systems that can be adapted to a wide variety of applications and environments. The SIFT analysis is then performed on two datasets of the same PUFs collected at two different times referred as PDLC1,2 for PDLC and (TiO2)1,2 TiO$_2$-PUF respectively. As expected, also for these PUFs, the number of matched features is maximized only for the 200 speckle patterns (R$_i$) produced from the same challenge (C$_i$) as indicated by the blue bars in Figure \ref{PDLCTiO2}b,e. Instead, for all the other cases, only a few points are matched, namely false positive, as indicated by the red bar. 
As a counterproof, the FHD analysis has been performed on these datasets, see Figure  \ref{PDLCTiO2}c,f. The PDLC presents a "like" distribution peaked at 0.12 and TiO$_2$ at 0.15, while the "unlike" distribution is centered at 0.5 for both PUFs which is in agreement with the results of PS-PUF.

Finally, we improved the SIFT algorithm to work with multiple-processors to parallelize its operations and make it even more suitable in real and industrial applications \cite{zhang2008sift, yang2020sift}. We performed a test comparing a selected speckle with a database of 1000 responses using workstations, laptops and a High Performance Computer HPC (the latter provided by Cineca). The results, reported in \textbf{Table \ref{table_time}}, demonstrate the speed of operation of the proposed SIFT method to recognize the target response in a large dataset. The time ranges from 18/19 sec for Apple Silicon M3 and M4 to 106 sec for HPC Leonoardo in single core while for multi-core operation, the time is dramatically reduced down to 5 sec using HPC to 30 sec using workstation CPUs Intel Xeon E5. This means that each single verification requires 5 microseconds, improving of 6 orders of magnitude the results reported in \cite{yeh2012robust}. This test proves the easiness in implementing the proposed method in industrial applications. 
\begin{table}[]
\label{table_time}
\begin{tabular}{|l|c|c|c|}
\hline
%\rowcolor[HTML]{C0C0C0} 
CPU model                                                                       & \begin{tabular}[c]{@{}c@{}}number of CPUs\\ RAM Gb\end{tabular} & \begin{tabular}[c]{@{}c@{}}Single Core \\ time (seconds)\end{tabular} & \begin{tabular}[c]{@{}c@{}}Multi Core \\ time (seconds)\end{tabular} \\ \hline
Intel Xeon E5                                                                   & \begin{tabular}[c]{@{}c@{}}4\\ 64\end{tabular}                 & 83                                                                    & 30                                                                   \\ \hline
%\rowcolor[HTML]{C0C0C0} 
Apple Silicon M1                                                                & \begin{tabular}[c]{@{}c@{}}8\\ 16\end{tabular}                 & 40                                                                    & 15                                                                   \\ \hline
Apple Silicon M3 - M4                                                                & \begin{tabular}[c]{@{}c@{}}11 - 10 \\ 18 - 16\end{tabular}                & 19 - 18                                                                     & 7 - 7                                                                    \\ \hline
%\rowcolor[HTML]{C0C0C0} 
Intel i7 10700                                                                       & \begin{tabular}[c]{@{}c@{}}8\\ 80\end{tabular}                  & 45                                                                    & 21                                                                    \\ \hline
\begin{tabular}[c]{@{}l@{}}Intel Xeon Platinum 8358\\ (HPC Leonardo)\end{tabular} & \begin{tabular}[c]{@{}c@{}}32\\ 512\end{tabular}               & 106                        & 5                       \\ \hline
\end{tabular}
\end{table}
%\newpage
\section{Conclusion}
We present a straightforward approach to analyze the uniqueness of speckle patterns by using image recognition SIFT algorithm for optical PUFs authentication. To benchmark its performance, we compared SIFT results with the Fractional Hamming Distance (FHD) method. Both methods were applied to various types of optical PUFs, including those made from dielectric nanospheres (polystyrene and TiO$_2$) and polymer-dispersed liquid crystals (PDLC). Our analysis highlights SIFT ability to extract and recognize several hundred unique features per speckle pattern. This capability even when comparing challenge-response pairs (CRPs) collected at different times and when identifying a target speckle pattern within a database. To demonstrate robustness, we applied the SIFT method to speckle images that were rotated, scaled, and cropped, achieving successful recognition in all cases. Additionally, the method supports real-time implementation as fast as 5 microseconds for each iteration using a multi-CPU setup. Overall, the proposed approach offers a fast, reliable, and scalable solution for both industrial and individual authentication systems, paving the way for effective anti-counterfeiting technologies based on optical PUFs.

\section*{Author contribution}
S.N fabricated and morphologically characterized the PUFs, G.E.L, F.R. and S.N settled up the experimental CRPs apparatus and acquire the data, G.E.L, M.D.L.B. and A.F. analyzed the data. G.E.L and A.F. conceived the idea and coordinated the overall research effort. The manuscript was written by G.E.L, M.D.L.B., S.N. and A.F with the input from all the authors.

\begin{acknowledgments}
G.E. Lio acknowledges the CINECA award (ID:HP10C1D8RJ) under the ISCRA initiative, for the availability of high performance computing resources and support. M.D.L.B, S.N. and A.F. acknowledge the financial support from project “PRIN 2022 2022T3B4HS-PE11 - Multi-step optical encoding in anticounterfeiting photonic tags based on liquid crystals (PHOTAG)” financed in the framework of Piano Nazionale di Ripresa e Resilienza (PNRR). F.R. acknowledges the project SERICS (PE00000014) under the MUR National Recovery and Resilience Plan funded by the European Union – NextGenerationEU and cofunded by the European Union - NextGeneration EU, "Integrated infrastructure initiative in Photonic and Quantum Sciences" - I-PHOQS [IR0000016, ID D2B8D520, CUP B53C22001750006] 
\end{acknowledgments}



\end{document}